\documentclass[10pt,a4paper,twocolumn]{vupreprint}
\usepackage[bindingoffset=0.5cm,textheight=25cm,hdivide={2.5cm,*,3cm}, vdivide={*,24cm,*}]{geometry}
\usepackage{amsmath}
\usepackage{amsfonts}
\usepackage{amssymb}
\usepackage[singlelinecheck=true,font=small]{caption}
\usepackage[numbers,square,comma,sort&compress]{natbib}
\usepackage{fancyhdr}
\usepackage{fancybox}
\usepackage{authblk}
\usepackage{ftnright}
\usepackage[pdftex]{graphicx}
\usepackage[pdftex,bookmarksnumbered=true,breaklinks=true]{hyperref}
\usepackage{subfig}
\usepackage{booktabs}
\newcommand{\eqnref}[1]{Eqn.~(\ref{#1})}		% for equations with preceding Eqn.
\newcommand{\figref}[1]{Fig.~\ref{#1}}			% for figures
\newcommand{\tabref}[1]{Tab.~\ref{#1}}			% for tables
\newcommand{\secref}[1]{Section~\ref{#1}}		% for sections
\newcommand{\ri}{{\rm i}}						% complex unit
\newcommand{\g}{\gamma}

\newcommand{\m}{\mu}

\newcommand{\s}{\sigma}

\renewcommand{\Xi}{\Xi}
\doi{10.1103/PhysRevD.91.102002}
\journalref{A. Almasi, Ph. Brax, D. Iannuzzi, R. I. P. Sedmik}{Phys. Rev. D}{91}{102002}{2015}{Force sensor for chameleon and Casimir force experiments with parallel-plate configuration}
\corrauthtext{{Corresponding author email: \url{a.almasi@vu.nl}}}

\author[1]{A. Almasi\footnote{xxx}}
\affil[1]{Department of Physics \& Astronomy and LaserLaB, VU university of Amsterdam (Netherlands)}
\author{Philippe Brax}

\affil[2]{Institut de Physique Th\'eorique, CEA, IPhT, CNRS, URA 2306, F-91191Gif/Yvette Cedex, France}
\author[1]{D. Iannuzzi}
\author[1]{R. I. P. Sedmik}

\hypersetup{pdfauthor={\theauthors},pdftitle={\thetitle}}
\graphicspath{{./}{./figures/}}
\begin{document}
	\twocolumn[\begin{@twocolumnfalse}%

		\maketitle

		%
		%%% VUPRINT: END NO-EDIT %%%
		%
		%=== CONTENT STARTS HERE ======================================================
		%
		\begin{abstract}
			The search for non-Newtonian forces has been pursued following many different paths. Recently it was suggested that hypothetical chameleon interactions, which might explain the mechanisms behind dark energy, could be detected in a high-precision force measurement. In such an experiment, interactions between parallel plates kept at constant separation could be measured as a function of the pressure of an ambient gas, thereby identifying chameleon interactions by their unique inverse dependence on the local mass density. During the past years we have been developing a new kind of setup complying with the high requirements of the proposed experiment. In this article we present the first and most important part of this setup -- the force sensor. We discuss its design, fabrication, and characterization. From the results of the latter we derive limits on chameleon interaction parameters that could be set by the forthcoming experiment. Finally, we describe the opportunity to use the same setup to measure Casimir forces at large surface separations with unprecedented accuracy, thereby potentially giving unambiguous answers to long standing open questions.
		\end{abstract}
		%
		%%% VUPREPRINT: DON'T EDIT THE FOLLOWING 2 LINES:
	\end{@twocolumnfalse}]
	\thisfancyput(-12pt,-719pt){\parbox{8cm}{\flushleft${}^\ast$\small\thecorrauthtext}}
%------------------------------------------------------------------------------
%Introduction}
%------------------------------------------------------------------------------
%
\section{Introduction}
\label{sec:Intro}
Over the last couple of decades, scientists have been proposing new experiments to measure the Casimir force between two objects kept at separations up to a few $\m$m that could give new insights on how the confinement of vacuum fluctuations modifies the attraction between two interacting surfaces~\cite{Rodriguez:2011,Klimchitskaya:2012}. The results reported in the literature have often been used to also set new limits on non-Newtonian forces and to explore if, at these separations, there is room for new physics~\cite{Decca:2007a,Lamoreaux:2012,Klimchitskaya:2012b}. Interestingly, apart from a few examples~\cite{Spaarnay:1958,Antonini:2009,Zou:2013}, all experiments to date have been carried out using either sphere-to-plate or crossed-cylinder geometries. The more obvious parallel-plate configuration can only be implemented if one can maintain a sufficient level of parallelism between the plates -- a technical hurdle that most experimentalists prefer to avoid. Still, the parallel-plate configuration allows measurements of surface interactions at much larger separations, providing new opportunities to tackle questions that cannot be answered unambiguously with curved-surface experiments~\cite{Brax:2007,Geraci:2008,Sushkov:2011a,Sushkov:2011}. One of the most interesting of these questions relates to the Khoury-Weltman theory~\cite{Khoury:2004}, which postulates that the accelerated expansion of the universe might be driven by a self-interacting scalar field, called \textit{chameleon}. The name is motivated by the peculiarity that the mass of this field depends on the density of matter in the local environment. If the chameleon field existed, it would manifest itself as an additional fifth force, which could be observable in Casimir-type force experiments. However, the relative magnitude of the chameleon interaction, with respect to other forces at the distances covered by current Casimir experiments, is too small to be detected. At larger distance (above $10\,\m$m), this relative magnitude increases, thereby providing better chances to detect chameleon interactions. In order to trigger strong enough forces to reach the detection limit in this distance regime, however, large interacting surfaces are required, which motivates the use of parallel plate configurations. In 2010, it has been shown~\cite{Brax:2010} that, if the chameleon theory is correct, a controlled change of the density of an ambient gas, in which two parallel plates of area $1\,{\rm cm}^2$ are kept at a separation of $10\,\m$m, could give rise to a change in the chameleon force of $\sim 1\, $pN. An experiment that could measure variations of $0.1\,$pN in the force between these plates could thus rigorously test the existence of chameleon fields. With a setup of this kind one could also perform accurate Casimir force measurements, giving one the possibility to finally settle a long standing debate on the correct description of the dielectric function of metals at zero frequency~\cite{Klimchitskaya:2012,Lamoreaux:2012}. \\
Inspired by this tantalizing opportunity, we have been developing a setup aiming to comply with all requirements to detect chameleon as well as Casimir forces with sub-pN precision between parallel plates at $\sim 10\,\m$m separation. The first step towards this Casimir and non-Newtonian force experiment (Cannex) is to demonstrate that one can fabricate a large area force sensor with the specified sensitivity. The goal of this paper is to show that this is indeed possible.\\
In \secref{sec:design} and \secref{sec:fabrication}, we introduce the mechanical design of the sensor and describe the processes that we have followed to fabricate a first prototype, respectively. In \secref{sec:character} we present the experimental characterization of this prototype, and compare the results to a numerical model of the design. In~\secref{sec:w_expect} and \secref{sec:casimir} we discuss the implications of the performance of our force sensor in the context of the upcoming Cannex experiment, and estimate the limits that could be set on chameleon parameters. Finally, \secref{sec:concl} summarizes our work.
%
%------------------------------------------------------------------------------
\section{Design}
\begin{figure}[!ht]
   \centering
   \includegraphics[width=82mm]{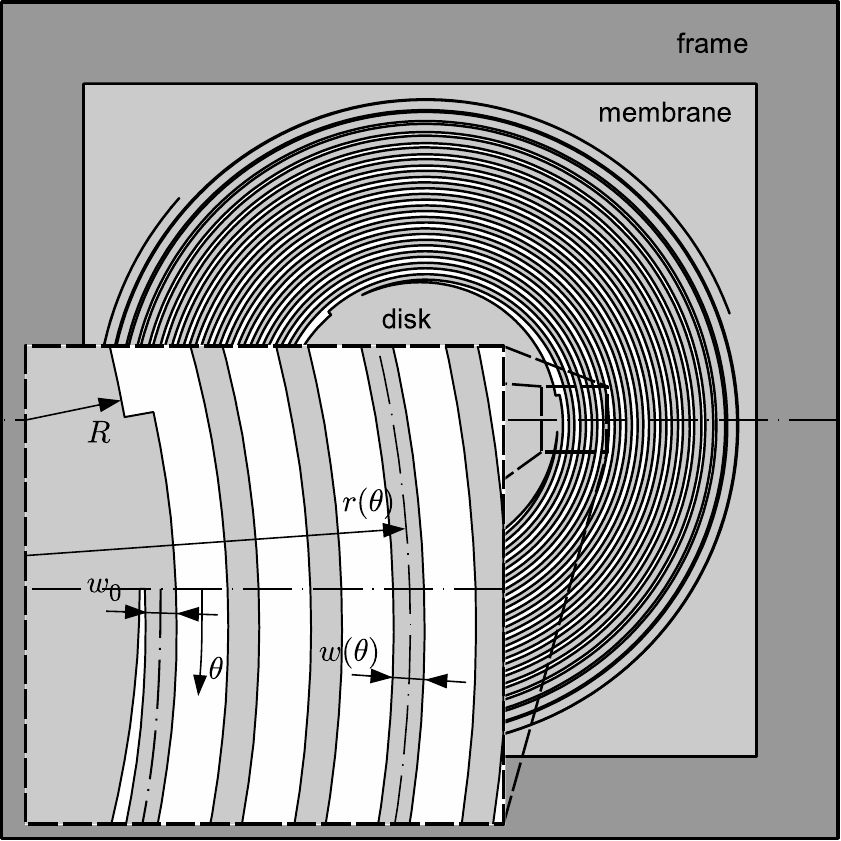}
   \caption{Top view of the force transducer. Tangential spring arms extend from a central disk onto which the measured forces act. Insert: enlarged view and geometric definitions. 
   	\label{fig:membrane}}
\end{figure}

%------------------------------------------------------------------------------
\label{sec:design}
Our force sensor consists of a spring-like structure that mechanically responds to forces orthogonally applied to a $1\,$cm$^2$ disk at its center. The resulting displacement is detected by measuring the capacitance between the disk and a fixed flat plate, placed parallel at a short separation. If the two plates rest at a separation of a few tens of $\m$m, one would measure a capacitance on the order of $100\,$pF. This value could be determined using a high precision capacitance bridge~\cite{AH2700manual} with a sensitivity of $0.1\,$ ppm. That level of precision would then translate into a displacement sensitivity on the order of $1\,$pm. In order to be able to detect force changes of $0.1\,$pN, the spring constant of the sensor should thus not exceed $0.1\,$N/m. A design complying with this requirement is shown in \figref{fig:membrane}. The geometry consists of three spiral-shaped arms extending tangentially from the central disk. The spacing of their center lines (dash-dotted curves in the figure) is even and described by the radius $r(\theta)=R+w_0/2+\theta N/(2\pi)(w_L+s)$, where $R$ is the radius of the disk, $s$ is the minimum spacing between arms, $N=3$ is the symmetry, $\theta\in [0,\theta_\text{max}]$ is the angle relative to the inner starting point, and $w_0=w(0)$ and $w_L=w(\theta_\text{max})$ are the widths at the inner and outer ends of the arm, respectively. In order to make optimal use of the available length and to achieve an approximately linear radial height profile (of the spring deformation), the width is chosen to follow a polynomial profile given by $w(\theta)=w_0+\left[f L(\theta)\right]^3$, where $f=(w_L-w_0)^{\frac{1}{3}}/L$, $L(\theta)$ is the length of the spiral at the position $\theta$, and $L$ is the total length. The geometric dimensions have been selected to minimize spacial requirements and the response to off-axes and tilt excitations, while maximizing stability and sensitivity.
%
%------------------------------------------------------------------------------
%Fabrication
%------------------------------------------------------------------------------
\section{Fabrication}
\label{sec:fabrication}
We have fabricated the force sensor described above following two different manufacturing techniques: wet etching of a silicon-on-insulator (SOI) wafer and laser cutting of a silicon wafer. 
The first method allows one to define the spring elements with lateral tolerances below $3\,\m$m, and results in perfectly clean surfaces. On the downside, it appears to have a high drop-out rate and proved to be costly. 
The second method is limited to larger lateral tolerances ($\sim10\,\m$m). Furthermore, the carving process exposes the sample to significant thermo-mechanical stress, which leads to crack formation at the edges of the mechanical structures, and contaminates the surfaces with back-sputtered debris, which has to be removed by immersing the devices in hydrogen fluoride (HF) solutions. The latter process leads to the formation of holes, eroded edges, and rougher surfaces.
Due to the high costs of production and the high drop-out rate, it was not possible to perform a systematic study of the devices fabricated via the wet etching procedure. Because this paper only aims at showing the overall feasibility of the force detection scheme, we have thus decided to focus on the second method. The results presented in this article have exclusively been obtained with laser-cut devices, using slightly different implementations (with respect to $\theta_\text{max}$ and $w_0$) of the geometry defined in \figref{fig:membrane}. For the sake of completeness, we note that, at the time of writing, a further production process using reactive ion etching is under development. The latter technique holds the promise for low tolerances and high quality surfaces at a comparably low cost. 
%
%------------------------------------------------------------------------------
%characterization
%------------------------------------------------------------------------------
\section{Mechanical characterization of the sensor}
\label{sec:character}
\subsection{Methods}
\label{sec:method}
\begin{figure} [!ht]
\centering
   \includegraphics[width=82mm]{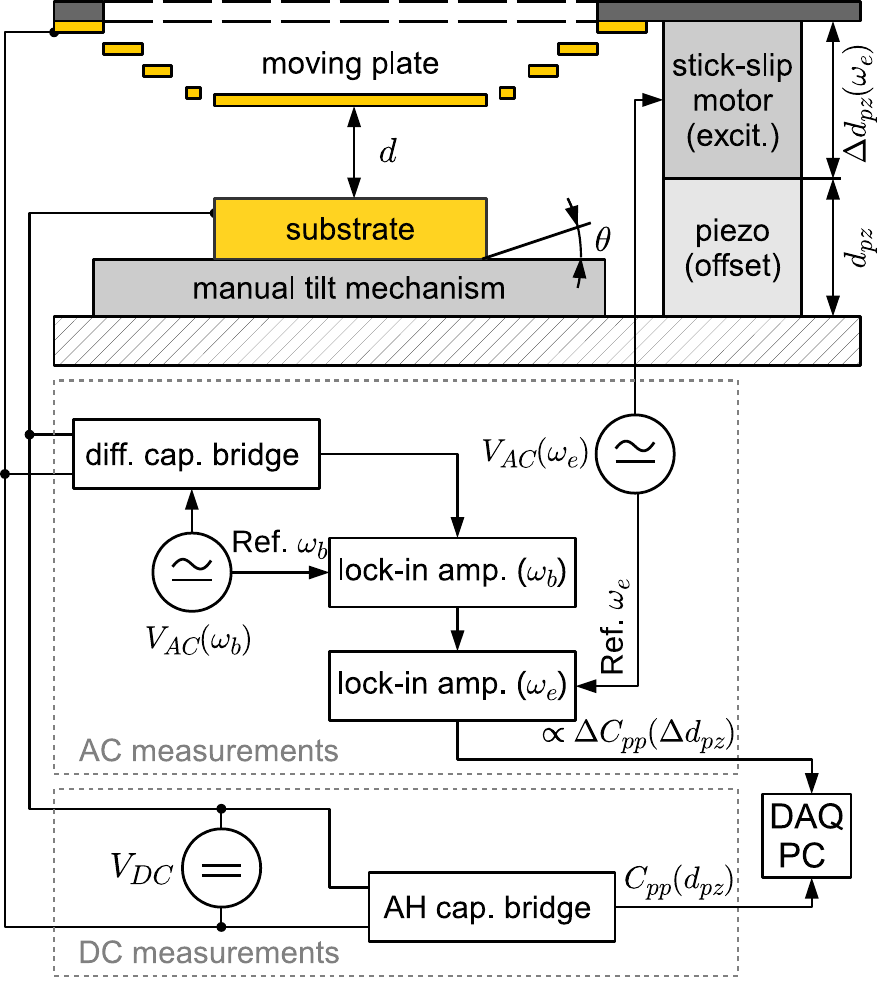}
\caption{Schematic view of the experimental setup. Note that the two different configurations for AC and DC measurements of the dynamic response and static deflections, respectively, are shown together.\label{fig:setup}}
\end{figure}
For the experimental characterization presented in this article we have constructed a simple setup, shown in \figref{fig:setup}.
 The sensing element described in \secref{sec:design} is mounted with its frame on top of a calibrated feedback-controlled piezo-electric translator (PI-P753, resolution $50\,$pm), which in turn is fixed to a stick-slip motor (Attocube-ANP101Z). The latter translator can be operated either in stick-slip mode for rough alignment ($d$ can be varied roughly from zero to $5\,$mm), or in piezo mode, which allows one to directly apply an external voltage $V_\text{AC}(\omega_e)$ to the piezo-electric transducer element to obtain continuous motion. Finally, the stack of actuators is mounted on a massive aluminum plate, which is thermally controlled to remain stable slightly above room temperature. A gold coated circular mica substrate of surface area $1\,$cm$^2$ is fixed with epoxy glue on top of a manual tilt mechanism (two-axis tilt adjustment plate for optical mirrors with an approximate range of $\pm10\,$deg.) and placed directly under the central disk of the sensor, thereby creating a parallel plate capacitance $C_\text{pp}$. The entire setup is placed inside a vacuum chamber (not shown) held at a pressure below $10^{-3}\,$mbar. As detailed below, two different detection circuits are utilized to monitor $d$ (and changes $\Delta d$ thereof) between the movable part of the transducer and the substrate.\\

First, we determine the dynamic mechanical response in an AC measurement by applying a small excitation $\Delta d_\text{pz}(\omega_e)=\Delta d_\text{pz}\cos(\omega_e t)$ to the sensor frame via the stick-slip motor operated in piezo mode~\footnote{Note that the driving voltage $V_\text{AC}(\omega_e)$ of the stick-slip piezo is generated directly by the internal generator of the utilized lock-in amplifier.}. The response amplitude $\Delta d$ of the sensor plate around the nominal fixed distance $d$ corresponds to a relative change $\Delta C_\text{pp}=\varepsilon_0 A T_{zz}(\omega_e)\Delta d_\text{pz}/d^2+\mathcal{O}\left(\delta d_\text{pz}^2/d^3\right)$, where,
\begin{equation}
T_{zz}(\omega)\equiv\frac{\Delta d}{\Delta d_\text{pz}}\approx\frac{\omega_0^2+\ri \omega\frac{\g_\text{eff}}{m_\text{eff}}}{\left(\omega_0^2-\omega^2+\ri \omega\frac{\g_\text{eff}}{m_\text{eff}}\right)}\,,
\label{eq:membrane_tfz_ver}
\end{equation}
is the transfer function from base vibration to absolute movement of the sensor plate. Note that we have assumed here that $T_{zz}(\omega)$ can be approximated by a simple one-dimensional mechanical oscillator with effective mass $m_\text{eff}$, damping coefficient $\g_\text{eff}$, and fundamental resonance frequency $\omega_0$. For amplitudes $\Delta d_{pz}=\mathcal{O}(1\,{\rm nm})$ and $d\approx 100\,\m$m, we have a signal amplitude $\Delta C_\text{pp}(\omega)/C_\text{pp}=\mathcal{O}\left(10^{-5}\right)$. In order to measure $\Delta C_\text{pp}(\omega)$ as function of frequency up to $\omega/(2 \pi)=100\,$Hz we use a differential analog capacitive bridge (General Radio $1616$) driven by an external supply voltage of amplitude $V_\text{AC}(\omega_b)=V_{\text{AC},b}\cos(\omega_b t)$ at frequency $\omega_b=2\pi\cdot 5\,{\rm kHz}\gg \omega_e$. Any variation $\Delta C_\text{pp}(t)$ results in an output $\propto \Delta C_\text{pp}(t)\cos(\omega_b t)$ of the bridge. In order to extract the net signal~\footnote{Both the differential bridge circuit and the mechanical system introduce additional phases $\varphi_b$ and $\varphi_e$, respectively. For the sake of clarity, these have been dropped here.} $\Delta C_\text{pp}(t)=\Delta C_\text{pp}\cos(\omega_e t)$, demodulation via a dedicated lock-in amplifier (Princeton research, P128-A) is required. The amplitude $\Delta C_\text{pp}$ can finally be measured by means of a second lock-in amplifier (Stanford research, SR830), referenced to $\omega_e$.\\
In DC measurements we detect slow variations ($<1\,$Hz) of the absolute value of $C_\text{pp}$ (and hence $d$) using an Andeen-Hagerling 2700A bridge, which for $d\lesssim 100\,\m$m ($C_\text{pp}\gtrsim9\,$pF) reaches $2.7\,$ppm resolution and $6.3\,$ppm accuracy. We perform static deflection measurements to demonstrate the principle of force detection with the sensor. In order to do so, a known fixed voltage $V_\text{DC}$ is applied between the sensor plate and the substrate, resulting in an attractive force $F_\text{es}(V_\text{DC})\approx (\varepsilon_0/2) A V_\text{DC}^2/(d-\Delta d)^2$. The resulting distance shift $\Delta d=d-d(V_\text{DC})$ corresponds to a change $\Delta C_{pp}=C_\text{pp}-C_\text{pp}(V_\text{DC})$, which can be measured conveniently for various settings of $d$ and $V_\text{DC}$.
%
%-----------------------------------------------------------------------------------
%result
%--------------------------------------------------------------------------------------
\subsection{Results}
\label{sec:result}
\begin{figure}  [!ht]
  \centering
  \includegraphics[width=82mm]{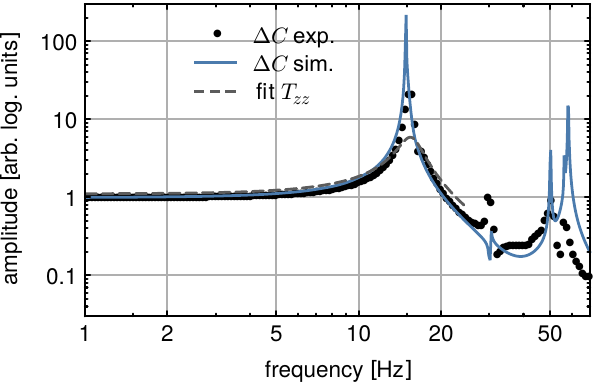}%fig_response_ac-eps-converted-to.pdf}
  \caption{Normalized dynamical response of the force transducer to mechanical excitation $\Delta d_\text{pz}$. A numerical simulation of the geometry using a constant base vibration amplitude (solid line) agrees very well with the experimental results (dots). The slight mismatch of $3\,$\% between the frequencies of the first resonance seen in experimental and numerical results ($15.20\pm0.15\,$Hz and $14.85\pm0.05\,$Hz, respectively) is a consequence of fabrication tolerances. Contrary to that, the simple model of \eqnref{eq:membrane_tfz_ver} (dashed line) fails to qualitatively predict the response.}
 \label{fig:results_ac}
\end{figure}

For the characterization of our design we have recorded the dynamic response $\Delta C_\text{pp}(\omega)$ as described above. In \figref{fig:results_ac} the acquired data are shown in normalized form and compared to a least squares fit to \eqnref{eq:membrane_tfz_ver} with free parameters $\omega_0$, $\g_\text{eff}$, and $m_\text{eff}$. Clearly, the model fails to predict the shape of the peak. It has been demonstrated in the literature~\cite{Erturk2008a} that effective (lumped parameter) models may not capture significant mechanical properties, even for simpler geometries than the one of our sensor. For this reason, we resorted to numerical 3D finite element computations using the geometry described in \secref{sec:design} with silicon as the material (assuming the elasticity tensor described in the literature~\cite{Hopcroft:2010} and a mass density of $2329\,$kg/m$^3$), and a vertical harmonic oscillation of the frame as excitation. As can be seen in \figref{fig:results_ac}, the numerical results agree very well with the measured response. Both the primary resonance at $15.20\pm0.15\,$Hz (representing a translational mode of the sensor plate) and the secondary resonance at $29.4\pm0.5\,$Hz (corresponding to a tilt mode) are matched by the simulation within $3\,$\%. The deviations in amplitude between numerical and experimental results for higher modes can be explained by considering several effects. First, tolerances of the laser cutting process create asymmetries and cross-couplings between different modes, which could not properly be taken into account in the numerical model. Second, the limited mechanical stability of the connection between the piezo stack and the transducer frame (see \figref{fig:setup}) results in a small angular tilt in addition to the vertical excitation, which amplifies the mode around $30\,$Hz. Finally, the spectral resolution of the measurement is insufficient to resolve all peaks properly. Being thus assured of the validity of the numerical results we computed the vertical displacement due to static forces, from which follows an elastic constant $k=0.63\pm 0.01\,$N/m. This value can be compared to an estimation $k_{\text{est}}\approx w_0^2 m_\text{eff}=0.30\pm0.01\,$N/m based on \eqnref{eq:membrane_tfz_ver}, where $m_\text{eff}\approx m_d+m_s/3$ is the effective sensor mass with contributions $m_d$ of the central disk and $m_s$ of the spring arms. The mismatch between $k$ and $k_{\text{est}}$ shows again that a one-dimensional effective model is not applicable in this case.

\begin{figure}  [!ht]
  \centering
  \includegraphics[width=82mm]{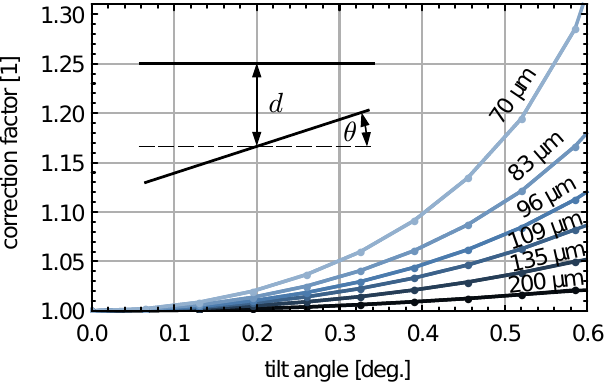}
  \caption{Correction factor $C_\text{pp}(d,\theta)/C_\text{pp}(d)$ due to the relative tilt angle $\theta$ between the plates for several separations $d$, as defined in the insert.}
 \label{fig:theta_d}
\end{figure}

\begin{figure}  [!ht]
	\centering
	    \includegraphics[width=82mm]{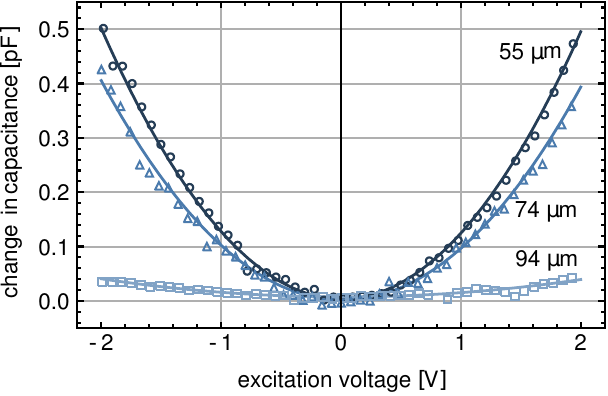}
	\caption{Measured $\Delta C_\text{pp}$ as a function of the applied $V_\text{DC}$ for three different nominal distances. The solid curves represent the best results obtained by least squares fits to the numeric model in Eqns.~\eqref{eq:model_delta_d}--\eqref{eq:model_delta_c}.\label{fig:results_dc}}
\end{figure}

Contrary to the dynamic measurements discussed above, where only the relative amplitude in dependence on the frequency was important, the experimental determination of the static response to forces acting onto the sensor's central plate demands for an accurate knowledge of the dependence of the capacitance on all involved parameters. Since in our test setup the angular alignment between the two plates is done manually, we have to take into consideration a residual relative tilt angle $\theta$ between them. The dependence $C_\text{pp}(d,\theta)$ can be estimated via dedicated computations using commercial finite element software (Comsol Multiphysics$^\text{\textregistered}$), leading to the results shown in \figref{fig:theta_d}.
Especially at short surface separations the change in $C_\text{pp}$ due to even small tilt angles is significant. While fringe fields contribute to errors at the percent level~\footnote{The measured capacitance will contain an offset due to fringe effects, which can be estimated~\cite{Hutson:1963,Chew:1980} to account for an error of $2.5\,$\% at $d=30\,\m$m and $0.6\,$\% at $d=10\,\m$m, respectively. An efficient way to reduce these effects is to use a ring-shaped Kelvin capacitance around the parallel plate configuration which is, however, beyond the scope of the current paper.}, the limited accuracy of the manual setting for $\theta$ in our test setup (estimated $0.5$ deg.) may yield changes in $C_\text{pp}$ of $10\,$\% or more.\\
In the static response measurements, electrostatic forces $F_\text{es}(d,\theta,V_\text{eff})=\big(V_\text{eff}^2/2\big) \partial C_\text{pp}(d,\theta)/\partial d$, with $V_\text{eff}=V_\text{DC}+V_e\cos(\omega_e t)+V_0$, is generated by applied voltages $V_\text{DC}$, the bridge excitation with amplitude $V_e$ (see \figref{fig:setup}), and the offset due to (local) intrinsic surface potentials $V_0$~\cite{Gaillard:2006}. Generally, such forces result in a change $\Delta d$ of $d$, and hence a variation $\Delta C_\text{pp}(d,\theta,V_\text{eff})$ in the capacity according to,
\begin{subequations}
\begin{align}
   k \Delta d &= F_\text{es}(d-\Delta d, \theta, V_\text{eff})\,,\label{eq:model_delta_d}\\
   C_\text{pp}(d,\theta,V_\text{eff})&=C_\text{pp}(d-\Delta d,\theta)\,,\label{eq:model_c}\\
\hspace*{-8pt}\text{and } \Delta C_\text{pp}(d,\theta,V_\text{eff})&= C_\text{pp}(d,\theta,V_\text{eff})-C_\text{pp}(d,\theta,0)\,.\label{eq:model_delta_c}
\end{align}
\end{subequations}
We interpolate the numerical results for $C_\text{pp}(d,\theta)$ shown in \figref{fig:theta_d} using a multi-dimensional spline fit to simultaneously compute $F_\text{es}(d-\Delta d,\theta,V_\text{eff})$ and to solve \eqnref{eq:model_delta_d} for $\Delta d$. This procedure yields a numerical parametric model which, again using spline interpolation, is amenable to extract $d$, $\theta$, and $V_0$ for known $V_{DC}$ and $V_e$ from experimental data on $C_\text{pp}(V_\text{DC})$ by performing least squares fits. 
%The obtained model is used to extract $d$, $\theta$, and $V_0$ for known $V_\text{DC}$ and $V_e$ from experimental data on $C_{\text{pp,exp}}(V_\text{DC})$ by performing least squares fits.
Numerical curves corresponding to the fitted parameters are shown together with experimental data for three different distances in \figref{fig:results_dc}. Explicit parameter data corresponding to these fits are given in \tabref{tab:fit1}.\\
\begin{table*}
  \centering
  \caption{Fitted parameters $d$, $\theta$, $V_0$, and the $\chi^2$ for the curves shown in \figref{fig:results_dc}. $\s$ gives the standard deviation of the fit residuals, which can be considered as a measure for noise.\label{tab:fit1}}	
\begin{tabular}{r@{.}l@{$\pm$}l r@{.}l@{$\pm$}l r@{$\pm$}l r@{.}l r@{.}l r@{.}l}
    \hline \hline
    \multicolumn{3}{c}{$d$ [$\mu m$]} & \multicolumn{3}{c}{$\theta$ [deg]} & \multicolumn{2}{c}{$V_0$ [mV]} & \multicolumn{2}{c}{$\chi^2$} & \multicolumn{2}{c}{$\s$ [fF]} & \multicolumn{2}{c}{$n_F$ [nN]}\\
    \hline
    93&8&2.6 & 0&45&0.10 & $-20$&29 & 0&922 & 3&7 & 5&1\\
    73&76&0.17 & 0&632&0.003 & $-14$&7 & 1&000 &10&9 & 14&9\\
    44&63&0.20 & 0&377&0.004 & $-6$&4 & 0&998 & 8&0 & 10&9\\
  \hline \hline
  \end{tabular}
\end{table*}
Note that the occurrence of three free parameters in the fits leads to large uncertainties, especially in $V_0$. For this reason, in the forthcoming Cannex experiment it will be of vital importance to actively reduce $\theta$ to below $1\,\m$rad via a feedback mechanism.

The sensitivity of our DC measurements can be estimated from the standard deviation $\s(\Delta C_\text{pp})$ of the normally distributed fit residuals given in \tabref{tab:fit1}. Noting that $\s(\Delta C_\text{pp})\approx(\varepsilon A/d^2)\sigma(d)$, and (for frequencies $\omega\ll \omega_0$), $\sigma(d)=\sigma(F)/k$, we can extract the corresponding force noise $n_F\approx\sigma(F)$ listed in \tabref{tab:fit1}. $n_F$ is approximately $5$ orders of magnitude larger than the targeted level for the Cannex experiment, which can intuitively be explained by the almost complete absence of isolation from electric and vibrational disturbances in the test setup in \figref{fig:setup}. This immediately demonstrates the necessity of an effective shielding and isolation system for the final experiment.
%-----------------------------------------------------------------------------------
% Expectations
%--------------------------------------------------------------------------------------
\section{Expected performance of the sensor in force measurements}
\label{sec:w_expect}
\begin{figure} [!ht]
 	\centering
 	\includegraphics[width=82mm]{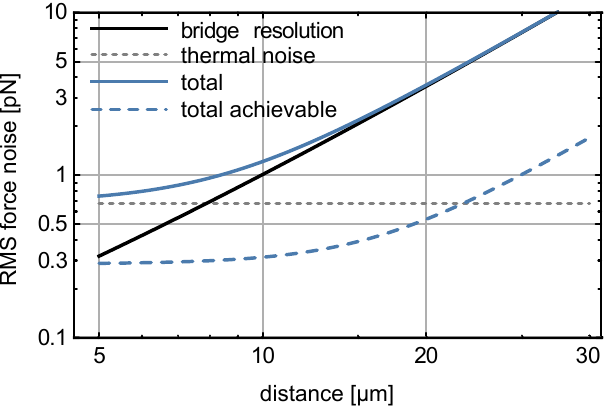}
 	\caption{Theoretical prediction of RMS noise levels in the final Cannex setup for a bridge detection bandwidth of $0.86\,$Hz~\cite{AH2700manual}. The thermal noise (dotted line) of laser-cut sensors and the bridge resolution (black line) are the main limiting factors for the achievable precision. Using an advanced version of the bridge and assuming a higher $Q$-factor for etched sensors leads to significantly better results (dashed line).\label{fig:limits}}
 \end{figure}

It is worthwhile to study the ultimate expected sensitivity limits imposed by the sensor design presented in this article. To start with, we assume the idealistic case that only the thermo-mechanical force noise $n_t$ of the sensor and the electronic noise $n_b$ of the bridge circuitry are present. Then we have~\cite{Gabrielson:1993},
\begin{align}
 n_t=\sqrt{\frac{4 k_B T \omega_0 m_\text{eff}}{Q}}\,,
\end{align}
where $Q$ is the quality factor of the sensor, $k_B$ is Boltzmann constant, and $T$ stands for temperature. From the simulation results, we estimate for the current sensor. Furthermore, we consider an effective bandwidth of our bridge circuit of $\Delta_\text{BW}=0.86\,$Hz~\cite{AH2700manual}, yielding an RMS noise amplitude $n_{t,\text{RMS}}=n_t\sqrt{\Delta_\text{BW}}\approx0.7\,$pN. Electronic noise $n_b$ presumably determines the resolution of the capacitance bridge. $n_{b}$ in units of capacity ($\Delta C_{\text{pp}}$) can be computed via an empirical model given by the manufacturer~\cite{AH2700manual}. Using $n_{bF}\approx k\Delta d$ and $\Delta d=d^2 n_b/\epsilon A$, we can rewrite $n_b$ in terms of force noise $n_{bF}$. For room temperature, $k=0.63\,$N/m and $V_e=0.3\,$Vpp, we obtain the results shown in \figref{fig:limits}. Assuming $n_t$ and $n_{bF}$ to be statistically uncorrelated, we compute the total RMS noise level for force measurements~\footnote{Note that the value $Q\approx 100$ used in the estimation is only valid for $P<10^{-3}\,$mbar.} as $n_\text{tot}=\sqrt{n_{t,\text{RMS}}^2+n_{bF}^2}\gtrsim 1.2\,{\rm pN}$ (light solid line). For comparison, we also include a model calculation assuming $Q=600$ and an advanced version (`option E') of the commercial bridge, yielding a much improved result (dashed line). The latter implies a sensitivity of better than $0.3\,{\rm pN}$ at plate separation $\leq10\,\m$m and can be regarded as the ideally achievable limit on the precision of Cannex in the current configuration. Further improvements could be made by altering the sensor design to allow for a larger $m_d$, and a higher $Q$-factor, or by using modulation techniques and a narrow detection bandwidth.\\
The mentioned limits regard the theoretically achievable precision with the sensor presented in this article. A full estimation of the accuracy would have to take into account the precise geometry, as well as residual fringe fields, parallelism, surface properties, external vibrations, electrical noise, and all occurring surface interactions. Yet, such an analysis is beyond the scope of the current paper. Based on the estimated resolution shown in \figref{fig:limits} we can, however, state expectations regarding new limits on hypothetical chameleon forces, which will be measured in a differential experiment where sensitivity is more important than accuracy.
 
\begin{figure} [!ht]
	\centering
	\includegraphics[width=82mm]{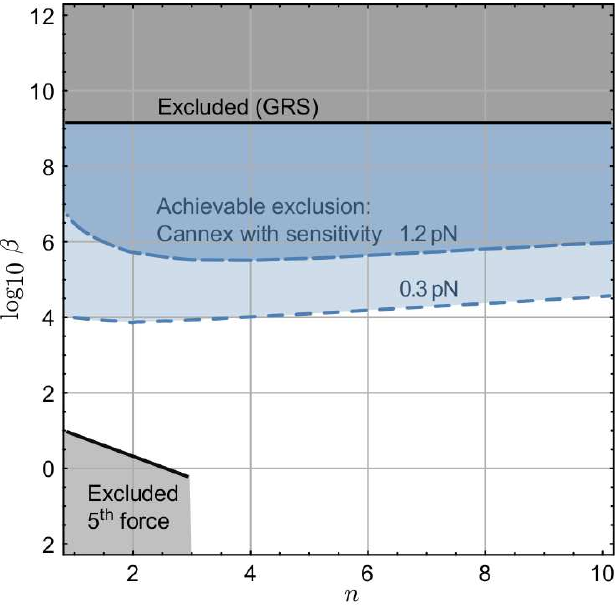}
	\caption{Limits on $\beta$ in dependence on the parameter $n$ defining the chameleon potential in \eqnref{eq:chameleonic}. Dashed lines present upper bounds achievable with our experiment for the sensitivity levels estimated in this article for measurements at $d=10\,\m$m. For comparison the best current limits found in the literature from gravity resonance spectroscopy~\cite{Jenke:2014} (top, GRS) and torsion balance fifth force measurements~\cite{Adelberger:2009} (bottom left, $5^{\rm th}$ force) are included.
		\label{fig:beta}}
\end{figure}

Two of us have previously estimated the expected chameleon force between two parallel plates immersed in a gaseous atmosphere of density $\rho$~\cite{Brax:2007,Brax:2010}. In this estimation, the authors have focused on chameleon potentials $V_{\phi}$ of the form,
\begin{equation}
V_{\phi}= \Lambda^{4}+\frac{\Lambda^{4+n}}{\phi^{n}}\,,
\label{eq:chameleonic}
\end{equation}
where $\phi$ is the chameleon field, and $\Lambda \simeq 2.4\times10^{-12}\,$GeV is chosen to match the cosmological constant associated with dark energy. The couplings of $\phi$ to all other matter fields~\cite{Khoury:2004} lead to an effective potential $V_{\text{E}}(\phi)=V(\phi)+\rho e^{\beta \phi/ m_{pl}}$, where $\beta$ and $m_{pl}$ are the (common) coupling constant and the reduced Planck mass, respectively. Interestingly $V_{\text{E}}$ has a local minimum $\phi_{\text{min}}$, which permits the definition of a mass $m^2_{\phi}=\left.\partial_{\phi\phi} V_{\text{E}}(\phi)\right|_{\phi_\text{min}}$ depending on $\rho$. As it has been shown in detail in~\cite{Brax:2007}, the equation of motion for $\phi$ can be solved for the parallel plate geometry. Denoting by $m_{b}$ and $\phi_b$, the mass and the field in the space between the plates, this solution can be written as an implicit relation between distance and field,
\begin{align}
d&=\frac{\sqrt{2} (z)^{(1+p)/2}}{m_b} \int_{0}^{1}\frac{x^{p-1} dx}{\sqrt{h_{p-1}(x)-z h_p(x)}}\,,\label{eq:chameleonic1}\\
\text{with } z&=\left(\frac{\phi_0}{\phi_b}\right)^{1/p}\!,\ p=(1+n)^{-1},\ \text{and } h_p(x)=\frac{1-x^p}{p}\,.\nonumber
\end{align}
Using \eqnref{eq:chameleonic1} one can then find from the gradient of $V_{\text{E}}(\phi)$ an expression for the chameleon pressure, 
\begin{align}
\frac{F_{\phi} (d,\rho)}{A}&=\frac{n+1}{n} \frac{\Lambda^{4+n}}{\phi_{b}^{n}} z^{1-p} \left[h_{1-p}(z)-z h_{-p}(z) \right], 
\label{eq:chameleonic2}
\end{align}
between the plates. $F_{\phi} (d,\rho)$ decreases with the local density $\rho$, while electrostatic, Casimir, and gravitational forces increase. This qualitative difference can be utilized to identify possible contributions of chameleon effects to the total measured force in the experiment. The amplitude $\Delta F_{\phi}=F_{\phi}(d,\rho_1)-F_{\phi}(d,\rho_2)$ of the difference in the chameleon force for two densities $\rho_1\neq\rho_2$ directly depends on the value of the parameter $\beta$. As proposed in \cite{Brax:2010}, such a variation in $\rho$ could be realized by immersing the plates in a gas of different pressures $P$. Hence, in a measurement of $\Delta F_{\phi}$ for different $P$, one could either see chameleonic effects, or set an upper limit for the coupling constant $\beta$, based on the force resolution of the experiment. \figref{fig:beta} shows constraints on $\beta$ as a function of on the parameter $n$ that would arise from measurements with xenon gas at room temperature at a plate separation of $10\,\m$m, and assuming the force resolutions estimated above. With a sensitivity of $1.2\,$pN, we could reduce the limit $\beta<5.8\times10^8$ given in Ref.~\cite{Jenke:2014} by approximately $3$ orders of magnitude. If the resolution of Cannex eventually reached $0.3\,$pN, a further improvement of $2$ orders of magnitude could be achieved. Hypothetically, with a sensitivity of $0.1\,$pN, one could test the complete range of $\beta$, and therefore, perform an absolute test for the existence of chameleon interactions.

\begin{figure} [!ht]
	\centering
	\includegraphics[width=82mm]{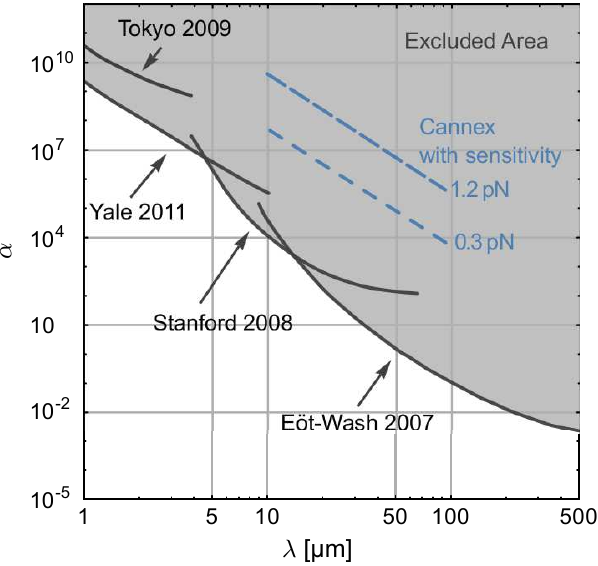}
	\caption{Limits on Yukawa-type forces arising from the estimated levels of sensitivity for the upcoming Cannex experiment (dashed lines). The shaded area marks parameter ranges already excluded by the results of experiments the E\"ot-Wash~\cite{Kapner:2007}, Stanford~\cite{Geraci:2008}, Yale~\cite{Sushkov:2011a}. and Tokyo~\cite{Masuda:2009} groups. 
		\label{fig:alpha}}
\end{figure}

By using $\alpha=2 \beta^2$ and $\lambda=m_{b}^{-1}$ in \eqnref{eq:chameleonic1} and \eqnref{eq:chameleonic2}, it is possible to express our limits on $\beta$ in the form of the familiar $\alpha$-$\lambda$ graph for Yukawa type (non-Newtonian) interactions, leading to the results shown in \figref{fig:alpha}. Note that the E\"ot-Wash~\cite{Kapner:2007} and Stanford~\cite{Geraci:2008} experiments, despite having a higher sensitivity for general Yukawa interactions, are not sensitive to chameleon forces due to the presence of an electrostatic shield~\cite{Brax:2007}.
%------------------------------------------------------------------------------
%Casimir
%------------------------------------------------------------------------------
\section{Casimir}
\label{sec:casimir}
The ability to perform force measurements with sub-pN precision at surface separations larger than $6\,\m$m yields the opportunity to not only detect non-Newtonian forces but also to possibly resolve a long standing question from a quite different field of physics - the Casimir effect.\\
For more than a decade (for recent reviews see~\cite{Klimchitskaya:2012,Lamoreaux:2012}) researchers have been trying to interpret experimental results in favor of either the Drude model $\varepsilon_D$ or the plasma model $\varepsilon_p$ as a valid description of the dielectric function of metals, but the available experimental data could not convince all researchers in favor of either candidates. One may write,
\begin{align}
 \varepsilon_D(\omega)=\varepsilon_{0}-\frac{\omega_p^2}{\omega^2+\ri\gamma\omega}\,,\quad\text{and } \varepsilon_p(\omega)=\varepsilon_{0}-\frac{\omega_p^2}{\omega^2}\,,
\end{align}
for the respective spectral dielectric functions, depending on the radial frequency $\omega$, and the plasma frequency $\omega_p=\sqrt{\rho_{e^-}e^2/m^*\varepsilon_{0}}$ of a metal with electrons of effective mass $m^*$, number density $\rho_{e^-}$, and elementary charge $e$, in a vacuum of dielectric constant $\varepsilon_{0}$. While both models are not rigorously defined from first principles, $\varepsilon_D$ appears to be more `physical' in the sense that it takes into account dissipation of electrons via the damping factor $\gamma$, which is neglected in $\varepsilon_p$. However, under certain conditions, the Drude model is thought to violate the second law of thermodynamics~\cite{Bezerra:2004,Geyer:2005}. If one uses the commonly applied Lifshitz theory to calculate the Casimir force $F_C$ between two plane parallel surfaces made of metals being described by either $\varepsilon_D$ or $\varepsilon_p$, the two results will differ by more than $100\,$\% at plate separations $d\gtrsim 6\,\mu$m~\cite{Bordag:2009}. For gold surfaces of area $1\,{\rm cm}^2$ the absolute value of $F_C$ at the same distance amounts to approximately $F_C\lesssim80\,$pN when assuming $\varepsilon_D$, and more than twice as much for $\varepsilon_p$. Hence, even if the final accuracy of Cannex cannot match its targeted precision level of $0.3\,$pN, a measurement of the Casimir force with absolute error of $10\,$\% appears to be well within the achievable range. Data produced by such an experiment in the distance regime $d>6\,\m$m would allow us to unambiguously discriminate if the Drude or the plasma model are in agreement with the experiment, and thereby, to settle a issue standing for more than ten years.
%
%------------------------------------------------------------------------------
%Conclusion}
%------------------------------------------------------------------------------
\section{Summary and conclusion}
\label{sec:concl}
The Casimir and non-Newtonian force experiment (Cannex) is poised to overcome the limitations of present precision force measurements by the utilization of macroscopic plane parallel plates, sub-pN force resolution, and an effective isolation system. In this paper we characterize the core element of this experiment -- the force sensor. This micro-machined device could be fabricated by wet etching, reactive ion etching, or laser cutting from silicon or SOI wafers. Geometrically, it consists of a flat central disk of area $1\,$cm$^2$, which is supported by spiral-shaped spring arms, allowing for a vertical translation of the disk. The design has been optimized to maximize the force sensitivity, while being robust enough for handling and to avoid instability under expected experimental conditions.\\
We have presented a static and dynamic characterization of the sensor structure. For this purpose a test setup was constructed in which the central disk is opposed to a fixed plate to form an electric capacitance. Using piezoelectric actuators we mechanically excited the sensor structure and monitored the vibration amplitude via a capacitive bridge circuit. We found that the mechanical response can not be efficiently described by a one-dimensional, lumped parameter model, but can be matched by a numerical simulation taking into account the precise geometry. By comparing numerical and experimental data we were able to determine the static force constant $k=0.63\,$N/m, and the quality factor $Q\approx100$ of a prototype device with larger-than-usual tolerances. In static measurements we could demonstrate the feasibility of force detection by applying an electric DC potential between the sensor and the fixed plate, resulting in a displacement which was again sensed capacitively. Due to the absence of electrical and vibrational insulation in our test setup, noise was found to be the limiting factor, leading to a force sensitivity of only $\sim 10\,$nN. Assuming that external influences (mechanical, electrical, and thermal) can sufficiently be attenuated, we estimated the residual Brownian noise of the sensor using the measured $Q$-factor and resonance frequency. Considering these results and the known electrical noise of the detection system we could eventually estimate the ideally achievable precision for Cannex to be around $1.2\,$pN (and $0.3\,$pN with straightforward improvements) at a plate separation of $\sim 10\,\m$m. This result gives confidence that the planned measurements can indeed unambiguously answer long-standing questions regarding the thermal contribution to the Casimir energy and give new limits to hypothetical chameleon forces, which could explain the nature of dark energy.\\
Before these measurements can be performed, however, we need to fully implement mechanisms to measure and maintain parallelism between our plates better than $1\,\m$rad, and to isolate our system from seismic, acoustic, thermal, and electric influences.
%
%------------------------------------------------------------------------------
% Acknowledgments
%------------------------------------------------------------------------------
\section{Acknowledgments}
This work was partially funded by the Foundation for Fundamental Research on Matter (FOM), which is financially supported by the Netherlands Organization for Scientific Research (NWO). R. Sedmik acknowledges his FWF Schr\"odinger fellowship J3050-N20 and donations by ASML and others to the Cannex crowdfunding initiative.
%===CONTENT ENDS HERE==========================================================

\begin{thebibliography}{31}%
	\makeatletter
	\providecommand \@ifxundefined [1]{%
		\@ifx{#1\undefined}
	}%
	\providecommand \@ifnum [1]{%
		\ifnum #1\expandafter \@firstoftwo
		\else \expandafter \@secondoftwo
		\fi
	}%
	\providecommand \@ifx [1]{%
		\ifx #1\expandafter \@firstoftwo
		\else \expandafter \@secondoftwo
		\fi
	}%
	\providecommand \natexlab [1]{#1}%
	\providecommand \enquote  [1]{``#1''}%
	\providecommand \bibnamefont  [1]{#1}%
	\providecommand \bibfnamefont [1]{#1}%
	\providecommand \citenamefont [1]{#1}%
	\providecommand \href@noop [0]{\@secondoftwo}%
	\providecommand \href [0]{\begingroup \@sanitize@url \@href}%
	\providecommand \@href[1]{\@@startlink{#1}\@@href}%
	\providecommand \@@href[1]{\endgroup#1\@@endlink}%
	\providecommand \@sanitize@url [0]{\catcode `\\12\catcode `\$12\catcode
		`\&12\catcode `\#12\catcode `\^12\catcode `\_12\catcode `\%12\relax}%
	\providecommand \@@startlink[1]{}%
	\providecommand \@@endlink[0]{}%
	\providecommand \url  [0]{\begingroup\@sanitize@url \@url }%
	\providecommand \@url [1]{\endgroup\@href {#1}{\urlprefix }}%
	\providecommand \urlprefix  [0]{URL }%
	\providecommand \Eprint [0]{\href }%
	\providecommand \doibase [0]{http://dx.doi.org/}%
	\providecommand \selectlanguage [0]{\@gobble}%
	\providecommand \bibinfo  [0]{\@secondoftwo}%
	\providecommand \bibfield  [0]{\@secondoftwo}%
	\providecommand \translation [1]{[#1]}%
	\providecommand \BibitemOpen [0]{}%
	\providecommand \bibitemStop [0]{}%
	\providecommand \bibitemNoStop [0]{.\EOS\space}%
	\providecommand \EOS [0]{\spacefactor3000\relax}%
	\providecommand \BibitemShut  [1]{\csname bibitem#1\endcsname}%
	\let\auto@bib@innerbib\@empty
	%</preamble>
	\bibitem [{\citenamefont {Rodriguez}\ \emph {et~al.}(2011)\citenamefont
		{Rodriguez}, \citenamefont {Capasso},\ and\ \citenamefont
		{Johnson}}]{Rodriguez:2011}%
	\BibitemOpen
	\bibfield  {author} {\bibinfo {author} {\bibfnamefont {A.~W.}\ \bibnamefont
			{Rodriguez}}, \bibinfo {author} {\bibfnamefont {F.}~\bibnamefont {Capasso}},
		\ and\ \bibinfo {author} {\bibfnamefont {S.~G.}\ \bibnamefont {Johnson}},\
	}\href
	{http://www.nature.com/nphoton/journal/v5/n4/full/nphoton.2011.39.html}
	{\bibfield  {journal} {\bibinfo  {journal} {Nature Phot.}\ }\textbf {\bibinfo
			{volume} {5}},\ \bibinfo {pages} {211} (\bibinfo {year}
		{2011})}\BibitemShut {NoStop}%
	\bibitem [{\citenamefont {Klimchitskaya}\ \emph
		{et~al.}(2012{\natexlab{a}})\citenamefont {Klimchitskaya}, \citenamefont
		{Bordag},\ and\ \citenamefont {Mostepanenko}}]{Klimchitskaya:2012}%
	\BibitemOpen
	\bibfield  {author} {\bibinfo {author} {\bibfnamefont {G.~L.}\ \bibnamefont
			{Klimchitskaya}}, \bibinfo {author} {\bibfnamefont {M.}~\bibnamefont
			{Bordag}}, \ and\ \bibinfo {author} {\bibfnamefont {V.~M.}\ \bibnamefont
			{Mostepanenko}},\ }\href
	{http://www.worldscientific.com/doi/abs/10.1142/S0217751X12600123} {\bibfield
		{journal} {\bibinfo  {journal} {Int. J. Mod. Phys. A}\
		}\textbf {\bibinfo {volume} {27}} (\bibinfo {year}
		{2012}{\natexlab{a}})}\BibitemShut {NoStop}%
	\bibitem [{\citenamefont {Decca}\ \emph {et~al.}(2007)\citenamefont {Decca},
		\citenamefont {L\'opez}, \citenamefont {Fischbach}, \citenamefont
		{Klimchitskaya}, \citenamefont {Krause},\ and\ \citenamefont
		{Mostepanenko}}]{Decca:2007a}%
	\BibitemOpen
	\bibfield  {author} {\bibinfo {author} {\bibfnamefont {R.~S.}\ \bibnamefont
			{Decca}}, \bibinfo {author} {\bibfnamefont {D.}~\bibnamefont {L\'opez}},
		\bibinfo {author} {\bibfnamefont {E.}~\bibnamefont {Fischbach}}, \bibinfo
		{author} {\bibfnamefont {G.~L.}\ \bibnamefont {Klimchitskaya}}, \bibinfo
		{author} {\bibfnamefont {D.~E.}\ \bibnamefont {Krause}}, \ and\ \bibinfo
		{author} {\bibfnamefont {V.~M.}\ \bibnamefont {Mostepanenko}},\ }
	\href {\doibase 10.1103/PhysRevD.75.077101} {\bibfield  {journal} {\bibinfo
			{journal} {Phys. Rev. D}\ }\textbf {\bibinfo {volume} {75}},\ \bibinfo
		{pages} {077101} (\bibinfo {year} {2007})}\BibitemShut {NoStop}%
	\bibitem [{\citenamefont {Lamoreaux}(2012)}]{Lamoreaux:2012}%
	\BibitemOpen
	\bibfield  {author} {\bibinfo {author} {\bibfnamefont {S.~K.}\ \bibnamefont
			{Lamoreaux}},\ }\href {\doibase
		10.1146/annurev-nucl-102711-095013} {\bibfield  {journal} {\bibinfo
			{journal} {Annu. Rev. Nucl. Part. Sci.}\ }\textbf {\bibinfo
			{volume} {62}},\ \bibinfo {pages} {37} (\bibinfo {year}
		{2012})}\BibitemShut {NoStop}%
	\bibitem [{\citenamefont {Klimchitskaya}\ \emph
		{et~al.}(2012{\natexlab{b}})\citenamefont {Klimchitskaya}, \citenamefont
		{Mohideen},\ and\ \citenamefont {Mostepanenko}}]{Klimchitskaya:2012b}%
	\BibitemOpen
	\bibfield  {author} {\bibinfo {author} {\bibfnamefont {G.~L.}\ \bibnamefont
			{Klimchitskaya}}, \bibinfo {author} {\bibfnamefont {U.}~\bibnamefont
			{Mohideen}}, \ and\ \bibinfo {author} {\bibfnamefont {V.~M.}\ \bibnamefont
			{Mostepanenko}},\ }\href {\doibase 10.1103/PhysRevD.86.065025} {\bibfield
		{journal} {\bibinfo  {journal} {Phys. Rev. D}\ }\textbf {\bibinfo {volume}
			{86}},\ \bibinfo {pages} {065025} (\bibinfo {year}
		{2012}{\natexlab{b}})}\BibitemShut {NoStop}%
	\bibitem [{\citenamefont {Spaarnay}(1958)}]{Spaarnay:1958}%
	\BibitemOpen
	\bibfield  {author} {\bibinfo {author} {\bibfnamefont {M.~J.}\ \bibnamefont
			{Spaarnay}},\ }\href@noop {}\href
	{http://www.sciencedirect.com/science/article/pii/S0031891458800907} {\bibfield
		{journal} {\bibinfo  {journal} {Physica}\ }\textbf {\bibinfo {volume} {24}},\
		\bibinfo {pages} {751} (\bibinfo {year} {1958})}\BibitemShut {NoStop}%
	\bibitem [{\citenamefont {Antonini}\ \emph {et~al.}(2009)\citenamefont
		{Antonini}, \citenamefont {Bimonte}, \citenamefont {Bressi}, \citenamefont
		{Carugno}, \citenamefont {Galeazzi}, \citenamefont {Messineo},\ and\
		\citenamefont {Ruoso}}]{Antonini:2009}%
	\BibitemOpen
	\bibfield  {author} {\bibinfo {author} {\bibfnamefont {P.}~\bibnamefont
			{Antonini}}, \bibinfo {author} {\bibfnamefont {G.}~\bibnamefont {Bimonte}},
		\bibinfo {author} {\bibfnamefont {G.}~\bibnamefont {Bressi}}, \bibinfo
		{author} {\bibfnamefont {G.}~\bibnamefont {Carugno}}, \bibinfo {author}
		{\bibfnamefont {G.}~\bibnamefont {Galeazzi}}, \bibinfo {author}
		{\bibfnamefont {G.}~\bibnamefont {Messineo}}, \ and\ \bibinfo {author}
		{\bibfnamefont {G.}~\bibnamefont {Ruoso}},\ }\href {\doibase 10.1088/1742-6596/161/1/012006}
	{\bibfield  {journal} {\bibinfo  {journal} {J. Phys.: Conf. Ser.}\ }\textbf
		{\bibinfo {volume} {161}},\ \bibinfo {pages} {012006} (\bibinfo {year}
		{2009})}\BibitemShut {NoStop}%
	\bibitem [{\citenamefont {{Zou}}\ \emph {et~al.}(2013)\citenamefont {{Zou}},
		\citenamefont {{Marcet}}, \citenamefont {{Rodriguez}}, \citenamefont
		{{Reid}}, \citenamefont {{McCauley}}, \citenamefont {{Kravchenko}},
		\citenamefont {{Lu}}, \citenamefont {{Bao}}, \citenamefont {{Johnson}},\ and\
		\citenamefont {{Chan}}}]{Zou:2013}%
	\BibitemOpen
	\bibfield  {author} {\bibinfo {author} {\bibfnamefont {J.}~\bibnamefont
			{{Zou}}}, \bibinfo {author} {\bibfnamefont {Z.}~\bibnamefont {{Marcet}}},
		\bibinfo {author} {\bibfnamefont {A.~W.}\ \bibnamefont {{Rodriguez}}},
		\bibinfo {author} {\bibfnamefont {M.~T.~H.}\ \bibnamefont {{Reid}}}, \bibinfo
		{author} {\bibfnamefont {A.~P.}\ \bibnamefont {{McCauley}}}, \bibinfo
		{author} {\bibfnamefont {I.~I.}\ \bibnamefont {{Kravchenko}}}, \bibinfo
		{author} {\bibfnamefont {T.}~\bibnamefont {{Lu}}}, \bibinfo {author}
		{\bibfnamefont {Y.}~\bibnamefont {{Bao}}}, \bibinfo {author} {\bibfnamefont
			{S.~G.}\ \bibnamefont {{Johnson}}}, \ and\ \bibinfo {author} {\bibfnamefont
			{H.~B.}\ \bibnamefont {{Chan}}},\ }\href
	{\doibase 10.1038/ncomms2842} {\bibfield  {journal} {\bibinfo  {journal}
			{Nature Comm.}\ }\textbf {\bibinfo {volume} {4}},\ \bibinfo {eid} {1845}
		(\bibinfo {year} {2013})}.\ 
	\bibitem [{\citenamefont {Brax}\ \emph {et~al.}(2007)\citenamefont {Brax},
		\citenamefont {van~de Bruck}, \citenamefont {Davis}, \citenamefont {Mota},\
		and\ \citenamefont {Shaw}}]{Brax:2007}%
	\BibitemOpen
	\bibfield  {author} {\bibinfo {author} {\bibfnamefont {Ph.}\ \bibnamefont
			{Brax}}, \bibinfo {author} {\bibfnamefont {C.}~\bibnamefont {van~de Bruck}},
		\bibinfo {author} {\bibfnamefont {A.~C.}\ \bibnamefont {Davis}}, \bibinfo
		{author} {\bibfnamefont {D.~F.}\ \bibnamefont {Mota}}, \ and\ \bibinfo
		{author} {\bibfnamefont {D.}~\bibnamefont {Shaw}},\ }\href {\doibase 10.1103/PhysRevD.76.124034} {\bibfield
		{journal} {\bibinfo  {journal} {Phys. Rev. D}\ }\textbf {\bibinfo {volume}
			{76}},\ \bibinfo {pages} {124034} (\bibinfo {year} {2007})}\BibitemShut
	{NoStop}%
	\bibitem [{\citenamefont {Geraci}\ \emph {et~al.}(2008)\citenamefont {Geraci},
		\citenamefont {Smullin}, \citenamefont {Weld}, \citenamefont {Chiaverini},\
		and\ \citenamefont {Kapitulnik}}]{Geraci:2008}%
	\BibitemOpen
	\bibfield  {author} {\bibinfo {author} {\bibfnamefont {A.~A.}\ \bibnamefont
			{Geraci}}, \bibinfo {author} {\bibfnamefont {S.~J.}\ \bibnamefont {Smullin}},
		\bibinfo {author} {\bibfnamefont {D.~M.}\ \bibnamefont {Weld}}, \bibinfo
		{author} {\bibfnamefont {J.}~\bibnamefont {Chiaverini}}, \ and\ \bibinfo
		{author} {\bibfnamefont {A.}~\bibnamefont {Kapitulnik}},\ }\href {\doibase 10.1103/PhysRevD.78.022002} {\bibfield
		{journal} {\bibinfo  {journal} {Phys. Rev. D}\ }\textbf {\bibinfo {volume}
			{78}},\ \bibinfo {pages} {022002} (\bibinfo {year} {2008})}\BibitemShut
	{NoStop}%
	\bibitem [{\citenamefont {Sushkov}\ \emph
		{et~al.}(2011{\natexlab{a}})\citenamefont {Sushkov}, \citenamefont {Kim},
		\citenamefont {Dalvit},\ and\ \citenamefont {Lamoreaux}}]{Sushkov:2011a}%
	\BibitemOpen
	\bibfield  {author} {\bibinfo {author} {\bibfnamefont {A.~O.}\ \bibnamefont
			{Sushkov}}, \bibinfo {author} {\bibfnamefont {W.~J.}\ \bibnamefont {Kim}},
		\bibinfo {author} {\bibfnamefont {D.~A.~R.}\ \bibnamefont {Dalvit}}, \ and\
		\bibinfo {author} {\bibfnamefont {S.~K.}\ \bibnamefont {Lamoreaux}},\
	}\href {\doibase
	10.1103/PhysRevLett.107.171101} {\bibfield  {journal} {\bibinfo  {journal}
		{Phys. Rev. Lett.}\ }\textbf {\bibinfo {volume} {107}},\ \bibinfo {pages}
	{171101} (\bibinfo {year} {2011}{\natexlab{a}})}\BibitemShut {NoStop}%
\bibitem [{\citenamefont {Sushkov}\ \emph
	{et~al.}(2011{\natexlab{b}})\citenamefont {Sushkov}, \citenamefont {Kim},
	\citenamefont {Dalvit},\ and\ \citenamefont {Lamoreaux}}]{Sushkov:2011}%
\BibitemOpen
\bibfield  {author} {\bibinfo {author} {\bibfnamefont {A.~O.}\ \bibnamefont
		{Sushkov}}, \bibinfo {author} {\bibfnamefont {W.~J.}\ \bibnamefont {Kim}},
	\bibinfo {author} {\bibfnamefont {D.~A.~R.}\ \bibnamefont {Dalvit}}, \ and\
	\bibinfo {author} {\bibfnamefont {S.~K.}\ \bibnamefont {Lamoreaux}},\
}\href {\doibase 10.1038/NPHYS1909} {\bibfield  {journal}
{\bibinfo  {journal} {Nature Phys.}\ }\textbf {\bibinfo {volume} {7}},\
\bibinfo {pages} {230} (\bibinfo {year} {2011}{\natexlab{b}})}\BibitemShut
{NoStop}%
\bibitem [{\citenamefont {Khoury}\ and\ \citenamefont
	{Weltman}(2004)}]{Khoury:2004}%
\BibitemOpen
\bibfield  {author} {\bibinfo {author} {\bibfnamefont {J.}~\bibnamefont
		{Khoury}}\ and\ \bibinfo {author} {\bibfnamefont {A.}~\bibnamefont
		{Weltman}},\ }\href {\doibase
	10.1103/PhysRevD.69.044026} {\bibfield  {journal} {\bibinfo  {journal}
		{Phys. Rev. D}\ }\textbf {\bibinfo {volume} {69}},\ \bibinfo {pages} {044026}
	(\bibinfo {year} {2004})}\BibitemShut {NoStop}%
\bibitem [{\citenamefont {Brax}\ \emph {et~al.}(2010)\citenamefont {Brax},
	\citenamefont {van~de Bruck}, \citenamefont {Davis}, \citenamefont {Shaw},\
	and\ \citenamefont {Iannuzzi}}]{Brax:2010}%
\BibitemOpen
\bibfield  {author} {\bibinfo {author} {\bibfnamefont {Ph.}\ \bibnamefont
		{Brax}}, \bibinfo {author} {\bibfnamefont {C.}~\bibnamefont {van~de Bruck}},
	\bibinfo {author} {\bibfnamefont {A.~C.}\ \bibnamefont {Davis}}, \bibinfo
	{author} {\bibfnamefont {D.~J.}\ \bibnamefont {Shaw}}, \ and\ \bibinfo
	{author} {\bibfnamefont {D.}~\bibnamefont {Iannuzzi}},\ }\href {\doibase 10.1103/PhysRevLett.104.241101}
{\bibfield  {journal} {\bibinfo  {journal} {Phys. Rev. Lett.}\ }\textbf
	{\bibinfo {volume} {104}},\ \bibinfo {pages} {241101} (\bibinfo {year}
	{2010})}\BibitemShut {NoStop}%
\bibitem [{\citenamefont {Andeen-Hagerling}(2014)}]{AH2700manual}%
\BibitemOpen
\bibfield  {author} {\bibinfo {author} {\bibnamefont {Andeen-Hagerling}},\
}\href {http://www.andeen-hagerling.com/ah2700a.htm} {\enquote {\bibinfo
	{title} {{AH 2700A} 50 hz - 20 khz ultra-precision capacitance bridge,
		operation and maintenance manual},}\ } (\bibinfo {year} {2014}),\ \bibinfo
{note} {online calculation sheet retrieved 2014-12-21}\BibitemShut {NoStop}%
\bibitem [{Note1()}]{Note1}%
\BibitemOpen
\bibinfo {note} {Note that the driving voltage $V_\protect \text {AC}(\omega
	_e)$ of the stick-slip piezo is generated directly by the internal generator
	of the utilized lock-in amplifier.}\BibitemShut {Stop}%
\bibitem [{Note2()}]{Note2}%
\BibitemOpen
\bibinfo {note} {Both the differential bridge circuit and the mechanical
	system introduce additional phases $\phi _b$ and $\phi _e$, respectively. For
	the sake of clarity, these have been dropped here.}\BibitemShut {Stop}%
\bibitem [{\citenamefont {Erturk}\ and\ \citenamefont
	{Inman}(2008)}]{Erturk2008a}%
\BibitemOpen
\bibfield  {author} {\bibinfo {author} {\bibfnamefont {A.}~\bibnamefont
		{Erturk}}\ and\ \bibinfo {author} {\bibfnamefont {D.J.}\ \bibnamefont
		{Inman}},\ }\href
{\doibase 10.1177/1045389X07085639} {\bibfield  {journal} {\bibinfo
		{journal} {J. Intell. Mater. Syst. Struct.}\ }\textbf
	{\bibinfo {volume} {19}},\ \bibinfo {pages} {1311} (\bibinfo {year}
	{2008})}\BibitemShut {NoStop}%
\bibitem [{\citenamefont {Hopcroft}\ \emph {et~al.}(2010)\citenamefont
	{Hopcroft}, \citenamefont {Nix},\ and\ \citenamefont
	{Kenny}}]{Hopcroft:2010}%
\BibitemOpen
\bibfield  {author} {\bibinfo {author} {\bibfnamefont {M.A.}\ \bibnamefont
		{Hopcroft}}, \bibinfo {author} {\bibfnamefont {W.D.}\ \bibnamefont {Nix}}, \
	and\ \bibinfo {author} {\bibfnamefont {T.W.}\ \bibnamefont {Kenny}},\
}\href {\doibase 10.1109/JMEMS.2009.2039697} {\bibfield
{journal} {\bibinfo  {journal} {J. Microelectromech. Syst.}\
}\textbf {\bibinfo {volume} {19}},\ \bibinfo {pages} {229} (\bibinfo
{year} {2010})}\BibitemShut {NoStop}%
\bibitem [{Note3()}]{Note3}%
\BibitemOpen
\bibinfo {note} {The measured capacitance will contain an offset due to
	fringe effects, which can be estimated~\cite {Hutson:1963,Chew:1980} to
	account for an error of $2.5\protect \tmspace +\thinmuskip {.1667em}$\% at
	$d=30\protect \tmspace +\thinmuskip {.1667em}\mu $m and $0.6\protect \tmspace
	+\thinmuskip {.1667em}$\% at $d=10\protect \tmspace +\thinmuskip {.1667em}\mu
	$m, respectively. An efficient way to reduce these effects is to use a
	ring-shaped Kelvin capacitance around the parallel plate configuration which
	is, however, beyond the scope of the current paper.}\BibitemShut {Stop}%
\bibitem [{\citenamefont {Gaillard}\ \emph {et~al.}(2006)\citenamefont
	{Gaillard}, \citenamefont {Gros-Jean}, \citenamefont {Mariolle},
	\citenamefont {Bertin},\ and\ \citenamefont {Bsiesy}}]{Gaillard:2006}%
\BibitemOpen
\bibfield  {author} {\bibinfo {author} {\bibfnamefont {N.}\ \bibnamefont
		{Gaillard}}, \bibinfo {author} {\bibfnamefont {M.}\ \bibnamefont
		{Gros-Jean}}, \bibinfo {author} {\bibfnamefont {D.}\ \bibnamefont
		{Mariolle}}, \bibinfo {author} {\bibfnamefont {F.}\ \bibnamefont
		{Bertin}}, \ and\ \bibinfo {author} {\bibfnamefont {A.}\ \bibnamefont
		{Bsiesy}},\ }\href {\doibase
        10.1063/1.2359297} {\bibfield  {journal} {\bibinfo
		{journal} {Appl. Phys. Lett.}\ }\textbf {\bibinfo {volume} {89}},\
	\bibinfo {pages} {154101} (\bibinfo {year} {2006})}\BibitemShut {NoStop}%
\bibitem [{\citenamefont {Gabrielson}(1993)}]{Gabrielson:1993}%
\BibitemOpen
\bibfield  {author} {\bibinfo {author} {\bibfnamefont {T.B.}\ \bibnamefont
		{Gabrielson}},\ }\href {\doibase 10.1109/16.210197} {\bibfield  {journal} {\bibinfo
		{journal} {IEEE Trans. Electron Devices}\ }\textbf {\bibinfo
		{volume} {40}},\ \bibinfo {pages} {903} (\bibinfo {year}
	{1993})}\BibitemShut {NoStop}%
\bibitem [{Note4()}]{Note4}%
\BibitemOpen
\bibinfo {note} {Note that the value $Q\approx 100$ used in the estimation is only valid for $P<10^{-3}\,$mbar.}\BibitemShut {Stop}%
\bibitem [{\citenamefont {Jenke}\ \emph {et~al.}(2014)\citenamefont {Jenke},
	\citenamefont {Cronenberg}, \citenamefont {Burgd\"orfer}, \citenamefont
	{Chizhova}, \citenamefont {Geltenbort}, \citenamefont {Ivanov}, \citenamefont
	{Lauer}, \citenamefont {Lins}, \citenamefont {Rotter}, \citenamefont {Saul},
	\citenamefont {Schmidt},\ and\ \citenamefont {Abele}}]{Jenke:2014}%
\BibitemOpen
\bibfield  {author} {\bibinfo {author} {\bibfnamefont {T.}~\bibnamefont
		{Jenke}}, \bibinfo {author} {\bibfnamefont {G.}~\bibnamefont {Cronenberg}},
	\bibinfo {author} {\bibfnamefont {J.}~\bibnamefont {Burgd\"orfer}}, \bibinfo
	{author} {\bibfnamefont {L.~A.}\ \bibnamefont {Chizhova}}, \bibinfo {author}
	{\bibfnamefont {P.}~\bibnamefont {Geltenbort}}, \bibinfo {author}
	{\bibfnamefont {A.N.}\ \bibnamefont {Ivanov}}, \bibinfo {author}
	{\bibfnamefont {T.}~\bibnamefont {Lauer}}, \bibinfo {author} {\bibfnamefont
		{T.}~\bibnamefont {Lins}}, \bibinfo {author} {\bibfnamefont {S.}~\bibnamefont
		{Rotter}}, \bibinfo {author} {\bibfnamefont {H.}~\bibnamefont {Saul}},
	\bibinfo {author} {\bibfnamefont {U.}~\bibnamefont {Schmidt}}, \ and\
	\bibinfo {author} {\bibfnamefont {H.}~\bibnamefont {Abele}},\ }\href {\doibase
	10.1103/PhysRevLett.112.151105} {\bibfield  {journal} {\bibinfo  {journal}
		{Phys. Rev. Lett.}\ }\textbf {\bibinfo {volume} {112}},\ \bibinfo {pages}
	{151105} (\bibinfo {year} {2014})}\BibitemShut {NoStop}%
\bibitem [{\citenamefont {Kapner}\ \emph {et~al.}(2007)\citenamefont {Kapner},
	\citenamefont {Cook}, \citenamefont {Adelberger}, \citenamefont {Gundlach},
	\citenamefont {Heckel}, \citenamefont {Hoyle},\ and\ \citenamefont
	{Swanson}}]{Kapner:2007}%
\BibitemOpen
\bibfield  {author} {\bibinfo {author} {\bibfnamefont {D.~J.}\ \bibnamefont
		{Kapner}}, \bibinfo {author} {\bibfnamefont {T.~S.}\ \bibnamefont {Cook}},
	\bibinfo {author} {\bibfnamefont {E.~G.}\ \bibnamefont {Adelberger}},
	\bibinfo {author} {\bibfnamefont {J.~H.}\ \bibnamefont {Gundlach}}, \bibinfo
	{author} {\bibfnamefont {B.~R.}\ \bibnamefont {Heckel}}, \bibinfo {author}
	{\bibfnamefont {C.~D.}\ \bibnamefont {Hoyle}}, \ and\ \bibinfo {author}
	{\bibfnamefont {H.~E.}\ \bibnamefont {Swanson}},\ }\href {\doibase
	10.1103/PhysRevLett.98.021101} {\bibfield  {journal} {\bibinfo  {journal}
		{Phys. Rev. Lett.}\ }\textbf {\bibinfo {volume} {98}},\ \bibinfo {pages}
	{021101} (\bibinfo {year} {2007})}\BibitemShut {NoStop}%
\bibitem [{\citenamefont {Adelberger}\ \emph {et~al.}(2009)\citenamefont
	{Adelberger}, \citenamefont {Gundlach}, \citenamefont {Heckel}, \citenamefont
	{Hoedl},\ and\ \citenamefont {Schlamminger}}]{Adelberger:2009}%
\BibitemOpen
\bibfield  {author} {\bibinfo {author} {\bibfnamefont {E.G.}\ \bibnamefont
		{Adelberger}}, \bibinfo {author} {\bibfnamefont {J.H.}\ \bibnamefont
		{Gundlach}}, \bibinfo {author} {\bibfnamefont {B.R.}\ \bibnamefont {Heckel}},
	\bibinfo {author} {\bibfnamefont {S.}~\bibnamefont {Hoedl}}, \ and\ \bibinfo
	{author} {\bibfnamefont {S.}~\bibnamefont {Schlamminger}},\ }\href {\doibase
	10.1016/j.ppnp.2008.08.002} {\bibfield  {journal} {\bibinfo
		{journal} {Prog. Part. Nucl. Phys.}\ }\textbf {\bibinfo
		{volume} {62}},\ \bibinfo {pages} {102} (\bibinfo {year}
	{2009})}\BibitemShut {NoStop}%
\bibitem [{\citenamefont {Masuda}\ and\ \citenamefont
	{Sasaki}(2009)}]{Masuda:2009}%
\BibitemOpen
\bibfield  {author} {\bibinfo {author} {\bibfnamefont {M.}~\bibnamefont
		{Masuda}}\ and\ \bibinfo {author} {\bibfnamefont {M.}~\bibnamefont
		{Sasaki}},\ }\href {\doibase
	10.1103/PhysRevLett.102.171101} {\bibfield  {journal} {\bibinfo  {journal}
		{Phys. Rev. Lett.}\ }\textbf {\bibinfo {volume} {102}},\ \bibinfo {pages}
	{171101} (\bibinfo {year} {2009})}\BibitemShut {NoStop}%
\bibitem [{\citenamefont {Bezerra}\ \emph {et~al.}(2004)\citenamefont
	{Bezerra}, \citenamefont {Klimchitskaya}, \citenamefont {Mostepanenko},\ and\
	\citenamefont {Romero}}]{Bezerra:2004}%
\BibitemOpen
\bibfield  {author} {\bibinfo {author} {\bibfnamefont {V.~B.}\ \bibnamefont
		{Bezerra}}, \bibinfo {author} {\bibfnamefont {G.~L.}\ \bibnamefont
		{Klimchitskaya}}, \bibinfo {author} {\bibfnamefont {V.~M.}\ \bibnamefont
		{Mostepanenko}}, \ and\ \bibinfo {author} {\bibfnamefont {C.}~\bibnamefont
		{Romero}},\ }\href {\doibase 10.1103/PhysRevA.69.022119} {\bibfield  {journal}
	{\bibinfo  {journal} {Phys. Rev. A}\ }\textbf {\bibinfo {volume} {69}},\
	\bibinfo {pages} {022119} (\bibinfo {year} {2004})}\BibitemShut {NoStop}%
\bibitem [{\citenamefont {Geyer}\ \emph {et~al.}(2005)\citenamefont {Geyer},
	\citenamefont {Klimchitskaya},\ and\ \citenamefont
	{Mostepanenko}}]{Geyer:2005}%
\BibitemOpen
\bibfield  {author} {\bibinfo {author} {\bibfnamefont {B.}~\bibnamefont
		{Geyer}}, \bibinfo {author} {\bibfnamefont {G.~L.}\ \bibnamefont
		{Klimchitskaya}}, \ and\ \bibinfo {author} {\bibfnamefont {V.~M.}\
		\bibnamefont {Mostepanenko}},\ }\href {\doibase 10.1103/PhysRevD.72.085009} {\bibfield
	{journal} {\bibinfo  {journal} {Phys. Rev. D}\ }\textbf {\bibinfo {volume}
		{72}},\ \bibinfo {pages} {085009} (\bibinfo {year} {2005})}\BibitemShut
{NoStop}%
\bibitem [{\citenamefont {Bordag}\ \emph {et~al.}(2009)\citenamefont {Bordag},
	\citenamefont {Fialkovsky}, \citenamefont {Gitman},\ and\ \citenamefont
	{Vassilevich}}]{Bordag:2009}%
\BibitemOpen
\bibfield  {author} {\bibinfo {author} {\bibfnamefont {M.}~\bibnamefont
		{Bordag}}, \bibinfo {author} {\bibfnamefont {I.~V.}\ \bibnamefont
		{Fialkovsky}}, \bibinfo {author} {\bibfnamefont {D.~M.}\ \bibnamefont
		{Gitman}}, \ and\ \bibinfo {author} {\bibfnamefont {D.~V.}\ \bibnamefont
		{Vassilevich}},\ }\href {\doibase 10.1103/PhysRevB.80.245406} {\bibfield  {journal}
	{\bibinfo  {journal} {Phys. Rev. B}\ }\textbf {\bibinfo {volume} {80}},\
	\bibinfo {pages} {245406} (\bibinfo {year} {2009})}\BibitemShut {NoStop}%
\bibitem [{\citenamefont {Hutson}(1963)}]{Hutson:1963}%
\BibitemOpen
\bibfield  {author} {\bibinfo {author} {\bibfnamefont {V.}~\bibnamefont
		{Hutson}},\ }\href {\doibase
	10.1017/S0305004100002152} {\bibfield  {journal} {\bibinfo  {journal}
		{Math. Proc. Cambridge Philos. Soc.}\ }\textbf
	{\bibinfo {volume} {59}},\ \bibinfo {pages} {211} (\bibinfo {year}
	{1963})}\BibitemShut {NoStop}%
\bibitem [{\citenamefont {Chew}\ and\ \citenamefont {Kong}(1980)}]{Chew:1980}%
\BibitemOpen
\bibfield  {author} {\bibinfo {author} {\bibfnamefont {Weng~Cho}\
		\bibnamefont {Chew}}\ and\ \bibinfo {author} {\bibfnamefont {Jin~Au}\
		\bibnamefont {Kong}},\ }\href {\doibase 10.1109/TMTT.1980.1130017} {\bibfield  {journal}
	{\bibinfo  {journal} {IEEE Trans. Microw. Theory Tech.}\
	}\textbf {\bibinfo {volume} {28}},\ \bibinfo {pages} {98} (\bibinfo
	{year} {1980})}\BibitemShut {NoStop}%
\end{thebibliography}
\end{document}